\documentstyle[aasms4,epsf]{article}

\begin{document}
\def\rpcomm#1{{\bf COMMENT by RP:  #1} \message{#1}}
\def\uptilde{\mathaccent"164}
\def\etal{{\it et al.~}} 
\def\ls{\vskip 12.045pt}   
\def\ni{\noindent}        
\def\kms{km\thinspace s$^{-1}$ }     
\def\amm{\AA\thinspace mm$^{-1}$ }     
\def\deg{\ifmmode^\circ _\cdot\else$^\circ _ \cdot$\fi }    
\def\degg{\ifmmode^\circ \else$^\circ $\fi } 
\def\solar{\ifmmode_{\mathord\odot}\else$_{\mathord\odot}$\fi} 
\def\arcs{\ifmmode {'' }\else $'' $\fi}     
\def\arcm{\ifmmode {' }\else $' $\fi}     
\def\buildrel#1\over#2{\mathrel{\mathop{\null#2}\limits^{#1}}}
\def\mper{\ifmmode \buildrel m\over . \else $\buildrel m\over .$\fi}
\def\hper{\ifmmode \rlap.^{h}\else $\rlap{.}^h$\fi}
\def\sper{\ifmmode \rlap.^{s}\else $\rlap{.}^s$\fi}
\def\arcsper{\ifmmode \rlap.{' }\else $\rlap{.}' $\fi}
\def\arcmper{\ifmmode \rlap.{'' }\else $\rlap{.}'' $\fi}
\def\gapprox{$_ >\atop{^\sim}$}     
\def\ltapprox{$_ <\atop{^\sim}$}     
\def\tworule{\noalign{\medskip\hrule\smallskip\hrule\medskip}}
\def\onerule{\noalign{\medskip\hrule\medskip}}
\def\et{{\it et~al.~}}
\newcommand{\lta}{{\small\raisebox{-0.6ex}{$\,\stackrel
{\raisebox{-.2ex}{$\textstyle <$}}{\sim}\,$}}}
\newcommand{\gta}{{\small\raisebox{-0.6ex}{$\,\stackrel
{\raisebox{-.2ex}{$\textstyle >$}}{\sim}\,$}}}
\newcommand{\apb}{$(A+B)/2$ }
\newcommand{\amb}{$(A-B)/2$ }

\def\apj{ApJ}  
\def\apjs{Ap.~J.~Suppl. }  
\def\apjl{Ap.~J.~ } 
\def\pasp{{Pub.~A.S.P.} }      
\def\mn{MNRAS}      
\def\aa{Astr.~Ap. }     
\def\aasup{AAS }     
\def\baas{Bull.~A.A.S. }  

\lefthead{Guti\'errez et al.}
\righthead{New CMB structures detected with the Tenerife experiments}

\title{NEW COSMOLOGICAL STRUCTURES ON MEDIUM ANGULAR SCALES DETECTED WITH THE TENERIFE EXPERIMENTS}

\author{C. M. Guti\'errez\altaffilmark{1}, S. Hancock\altaffilmark{2}, R. D. Davies\altaffilmark{3}, R. Rebolo\altaffilmark{1}, R. A. Watson\altaffilmark{1,3}, R. J. Hoyland\altaffilmark{1},
A. N. Lasenby\altaffilmark{2}, and A. W. Jones\altaffilmark{2}}

\altaffiltext{1}{Instituto de Astrof\'\i sica de Canarias, 38200 La Laguna, Tenerife, SPAIN}

\altaffiltext{2}{Mullard Radio Astronomy Observatory, Cavendish Laboratory, 
        Madingley Road, Cambridge CB3 OHE, UK}

\altaffiltext{3}{University of Manchester, Nuffield Radio Astronomy 
        Laboratories, Jodrell Bank, Macclesfield, Cheshire SK11 9DL, UK}

\begin{abstract}

We present observations at 10 and 15 GHz taken with the Tenerife experiments
in a band of the sky at Dec.=+35\degg. These experiments are sensitive
to multipoles in the range $l=10-30$. The sensitivity per beam is 56
and 20 $\mu$K for the 10 and the 15 GHz data, respectively. After subtraction of the prediction of known radio-sources, the analysis
of the data at 15 GHz at high Galactic latitude shows the presence of a
signal with amplitude $\Delta T_{\rm RMS}\sim 32$ $\mu$K. In the case of a
Harrison-Zeldovich spectrum for the primordial fluctuations, a
likelihood analysis shows that this signal corresponds to a quadrupole amplitude $Q_{\rm RMS-PS}=20.1^{+7.1}_{-5.4}$ $\mu$K, in agreement with our
previous results at Dec.+=40\degg~and with the results of the COBE DMR. There is clear
evidence for the presence of individual features in the RA range
190\degg~to 250\degg~with a peak to peak amplitude of $\sim$110 $\mu$K. A
preliminary comparison between our results and COBE DMR predictions for
the Tenerife experiments clearly indicates the presence of individual
features common to both. The constancy in amplitude over such a large
range in frequency ($10-90$ GHz) is strongly indicative of an intrinsic
cosmological origin for these structures.

\end{abstract}

\keywords{cosmology~-~large scale structure of the universe~-~cosmic microwave background}

\section{INTRODUCTION}

The study of the cosmic microwave backround (CMB) has produced crucial
advances in cosmology in recent years with the statistical detection of
the fluctuations by the COBE DMR instrument (Smoot \et 1992; Bennett
\et 1996), the direct observation of CMB features  on scales $\gta 5$\degg~(Hancock \et 1994) and the detections of signals on smaller angular
scales down to a few arcminutes (see Hancock \et 1997$a$ for a summary
of the present observational status). Current observations permit
the determination of the overall level of normalization and constrain
the CMB power spectrum on multipoles up to $l\sim 700$ thereby
providing good evidence for the existence of the Doppler peak predicted
by standard inflationary scenarios (Hancock \et 1997$a$). Potentially,
the power spectrum of the CMB fluctuations will allow the cosmological
parameters (the curvature of the Universe, the baryonic content, the
Hubble constant, etc) to be determined to an uncertainty of a few
per cent.  Additionally, the possible existence of a foreground of
gravitational waves (Crittenden \et 1993; Steinhardt 1994) can be determined from measurements of the CMB power spectrum on angular scales $\gta
2\degg$ ($l\lta 30$). To achieve these goals, more sensitive observations are needed over a wide range of angular scales and covering larger regions of the sky; observations over a range of frequencies are required to identify and remove Galactic foregrounds. The results presented here correspond to scales 2-10\degg~and
therefore can help to establish the overall level of normalization and
test for the presence of gravitational waves. In the following, a description of the instrumental set-up and the observations (Section 2) is followed by an analysis of the data including a joint-likelihood analysis with the previous results of our experiment at an adjacent declination (Section 3). Our conclusions are given in Section 4.

\section{OBSERVATIONS}

The Tenerife CMB experiments have been extensively described in
previous papers (see for instance Davies \et 1996). Basically the suite
of instruments consists of three radiometers, each with two independent
channels, operating at frequencies of 10, 15, and 33 GHz. The
instruments use a double-differencing technique where one of the
resultant three 5\degg~beams lies on the meridian and the other two
(negative and half the amplitude of the central beam) are separated  by
$\sim 8\degg$ in right ascension. Each day the instruments scan a band
of the sky at constant declination. Observations are repeated over many days in
order to achieve the sensitivity levels necessary to observe CMB
anisotropies. Data processing includes the removal of points taken in
poor atmospheric conditions, or corresponding to times when technical
failures in the instrumental system have occurred. Also all data taken
closer than 50\degg~to the Sun or 30\degg~to the Moon are removed. The
remaining data are dominated by low-order baseline variations which
differ from day to day and have an atmospheric origin (Davies \et
1992, 1996).  The angular scales of such variations are larger than the beam
response of the instrument and are removed by a maximum entropy method
(Jones \et 1997). Results published up to now include observations at
10, 15, and 33 GHz in a single strip at Dec.=+40\degg~(Watson \et 1992; 
Hancock \et 1994) along with data at several declinations  ranging from
+35\degg~to +45\degg~at 10 GHz and from 37.5\degg~to 42.5\degg~at 15
GHz (Guti\'errez \et 1995).

The data presented here cover a band of the sky at Dec.=+35\degg~and
were taken in several observing campaigns between 1988 March and 1996 April
at 10 GHz, and between 1992 March and 1995 March at 15 GHz. In the
region of the scan at high Galactic latitude the number of independent
observations contributing to each 1\degg~bin in RA is between 50 and
110 at 10 GHz, and between 60 and 160 at 15 GHz depending on the RA,
with more observations at higher RA. The mean sensitivities in a
beam-sized region (5\degg$\times$5\degg) are 56 and 20 $\mu$K at 10 and
15 GHz, respectively. This represents an improvement by more than a
factor of 2 with respect to the previous results (Guti\'errez \et 1995)
at 10 GHz. Figure 1 shows the stacked results in the region
RA=$160\degg-250\degg$ corresponding to Galactic latitudes $b\gta
40$\degg. Data are shown binned in a 4\degg~interval in RA and
therefore a single structure on the sky is sampled with at least 6
points across the triple-beam response. The maximum excursion at both frequencies is $\sim 100$ $\mu$K. The
sensitivity of the data at 15 GHz is sufficiently high to show
clear  structures in the range RA=190\degg$-$250\degg. The statistical
amplitude of the signals is quantified in the following sections.

\section{ANALYSIS}

\subsection{The Foregrounds}

We have concentrated our analysis on the region RA=$161\degg-250\degg$, 
which is at Galactic latitude $b\gta 40$\degg. In this
section we consider the possible sources of non-CMB foregrounds in our data. The
contribution from known point sources has been calculated using the
Kuhr \et (1981) catalogue and the Green Bank sky survey (Condon, 
Broderick \& Seielstad 1989) complemented by the Michigan and Metsahovi monitoring
programme. In the Dec.=+35\degg~band of the sky the most intense
radio-source at high Galactic latitude is 1611+34
(RA=16$^{\rm h}$11$^{\rm m}$48$^{\rm s}$, Dec.=34\degg$20'20''$) with a flux density $\sim
2$ Jy at both 10 and 15 GHz. After convolution with the triple beam of
our experiment this source gives peak amplitudes of $\sim 60$ and $\sim
$30 $\mu$K at 10 and 15 GHz, respectively, which are at the limit of
detection of each data set.  The rms point
source contributions along the Dec.=+35\degg~scan are 26 and 11 $\mu$K
at 10 and 15 GHz, respectively. These values agree closely with 27 and
12 $\mu$K at 10 and 15 GHz read directly from the plots of Franceschini
\et (1989) for a beamwidth of 5\degg. The agreement is not surprising
since the main contribution to the point-source background is from
sources stronger than 0.1 Jy, all of which are included in the Green
Bank survey. Point sources are therefore only responsible for a small
fraction of the detected signals. In the following analysis we have
subtracted this point-source contribution.

In previous papers (Davies, Watson \&
Guti\'errez 1996), we have demonstrated the unreliability of the high-frequency predictions of the diffuse Galactic foreground  using the low-frequency
surveys at 408 MHz (Haslam \et 1982) and 1420 MHz (Reich \& Reich 1988) and have emphasized the necessity of using observations at higher frequencies.
It is possible to estimate the magnitude of such a
contribution from a comparison between our own measurements at 10 and
15 GHz and the COBE DMR results at 31, 53, and 90 GHz. In Sections 3.2 and 3.3 it will be shown how the Galactic foreground can contribute only with a small fraction to the signal detected at 15 GHz.

\subsection{Estimates of the CMB Fluctuation Level}
 
Figure 2 presents the auto-correlation of the 15-GHz data in the region
at RA=$161\degg-250\degg$. The error-bars were determined by standard
Monte Carlo techniques. These techniques were also used to obtain the
confidence bands in the case of pure uncorrelated noise (long-dashed
lines) and the expected correlation (short-dashed line) in the case of
a Harrison-Zeldovich spectrum for the primordial fluctuations with an
amplitude corresponding to the signal of maximum likelihood (see next
paragraph). Clearly the Harrison-Zeldovich model gives an adequate
description of the observed correlation and shows that the results are
incompatible with pure uncorrelated noise. The cross-correlation
between the data at 10 and 15 GHz is inconclusive as it is
dominated by the noisy character of the 10-GHz data.

We have applied a likelihood analysis described in detail by Hancock
\et (1994) to the data at 10 and 15 GHz in the range
RA=$161\degg-250\degg$. Assuming a Harrison-Zeldovich spectrum for the
primordial fluctuations, the likelihood curve for the 15-GHz data shows
a clear peak  ($5.5\times $10$^4$ normalized respect to the value for
zero signal) at a rms temperature fluctuation in the data $\Delta
T_{\rm RMS}\sim 32$ $\mu$K. Analyzing the curve in a {\it Bayesian}
sense with uniform prior we obtained $\Delta T_{\rm RMS}=32^{+11}_{-9}$
$\mu$K, which corresponds to an expected power-spectrum normalization
with a quadrupole amplitude $Q_{\rm RMS-PS}=20.1^{+7.1}_{-5.4}$ $\mu$K
(68\% confidence level). These results do not depend strongly on the
precise region analyzed; for instance analyzing the sections
RA=$161\degg-230\degg$~or RA=$181\degg-250\degg$, we obtain $Q_{\rm
RMS-PS}=19.0^{+9.0}_{-6.5}$ $\mu$K and $Q_{\rm
RMS-PS}=20.0^{+8.0}_{-6.0}$ $\mu$K, respectively. By using Monte Carlo techniques (5000 simulations) we have demonstrated that, for our data, the maximum likelihood estimator is essentially unbiased. As a consequence of
the noise level in the 10-GHz data, there is no evidence of signal in
the likelihood analysis of the data at this frequency. However, we
obtained a limit on $Q_{\rm RMS-PS}\le 33.8$ $\mu$K (95\% C.L.), which
is compatible with the amplitude of the signal detected at 15 GHz. The
signal at 15 GHz ($Q_{\rm RMS-PS}=20.1$ $\mu$K) is compatible with the
signal $Q_{\rm RMS-PS}=18$ $\mu$K present in the COBE DMR data (Bennett
\et 1996) also estimated on the Harrison-Zeldovich model. However, assuming that
COBE DMR data give the correct normalization for the cosmological signal,
our slightly higher normalization may be a signature of some small galactic contamination in the data at 15-GHz.

We have run a joint likelihood analysis (Guti\'errez \et 1995) between
the 15-GHz data presented here and those at Dec.=40\degg~presented in
Hancock \et (1994) taking into account the small corrections due to the
presence of atmospheric correlated noise between channels (Davies \et
1996). The angular separation between the strips is equal to the
half-power beamwidth of the individual antennas and corresponds to an
overlap of $\sim 30$\% of the beam areas. As a consequence the scans at
Dec.=+35\degg~and Dec.=+40\degg~are largely independent. We have chosen
the region between RA=$161\degg-230\degg$~at Dec.=+40\degg~and thus
excluded the variable radio source 3C 345 (RA$\sim 250\degg$) that was
clearly detected in our data. The combined data set was used to obtain
quadrupole amplitude for different assumed values of the spectral index
$n$ ($P(k)\propto k^n$). Figure 3 shows the amplitude of the signal for
each model and the one-sigma level bounds in the spectral index versus
$Q_{\rm RMS-PS}$ plane. The relation between the expected quadrupole
and the spectral index in such tilted models can be parameterized by
$Q_{\rm RMS-PS}=25.8^{+8.0}_{-6.5}\exp\{-(n-0.81)\}$ $\mu$K. In the
case of a flat spectrum ($n=1$) we obtain $Q_{\rm
RMS-PS}=21.0^{+6.5}_{-5.5}$ $\mu$K, in agreement with the results
obtained analyzing independently the strips at Dec.=+35\degg~(see
above) and Dec.=+40\degg~($Q_{\rm RMS-PS}=22^{+10}_{-6}$ $\mu$K,
Hancock \et 1997$b$). Our data can not be used on their own to
constrain effectively the spectral index $n$,  but we have already
established limits on $n$ from a comparison of our Dec.=+40\degg~data
and the COBE DMR results (see the discussion in Hancock \et 1997$b$). A
refinement of these estimates by including the Dec.=+35\degg~data will
be discussed in a forthcoming paper.

\subsection{Comparison with Features in the COBE DMR Data}

A first direct comparison of the Tenerife and COBE DMR data was made by
Lineweaver \et (1995) who demonstrated a clear correlation between the
data sets at Dec.=+40\degg~and showed the presence of common individual
features. Bunn, Hoffman \& Silk (1996) applied a Wiener filter to the
two-year COBE DMR data assuming a CDM model. They obtained a weighted
addition of the results at the two more sensitive frequencies (53 and
90 GHz) in the 7\degg~beam COBE DMR data, and used the results of this
filtering to compute the prediction for the Tenerife beam-switching
experiments over the region $35\degg \le$Dec.$\le 45\degg$. At high
Galactic latitude the most significant features predicted for the
Tenerife data are two hot spots with peak amplitudes $\sim 50$-$100$
$\mu$K around Dec.=+35\degg~at RA$\sim 220$\degg~and $\sim250\degg$.
Figure 4 compares the Bunn \et (1996) prediction with the maximum
entropy reconstruction (Jones \et 1997) of our 15-GHz data reconvolved
to be consistent with our beam geometry. The solid line shows the
reconvolved results at 15 GHz after subtraction of the known
point-source contribution as described in Section 3.1. The two most
intense structures in these data agree in amplitude and position with
the predictions from 53 and 90 GHz (dashed line), with only a slight
shift in position for the feature at RA=250\degg. A possible
uncertainty by a factor as large as 2 in the contribution of the point
source 1611+34 would change the shape and amplitude of this second
feature only slightly. In the range RA=160\degg-200\degg, there are more
structure in the Tenerife reconstruction as compared with the predictions from
the COBE DMR data. In that region, the Tenerife data are less sensitive
as compared with the observations at higher RA, and therefore the
maximum entropy reconstruction is less reliable. Other possible reasons
for the discrepancy between both data-sets in that region, could be due
to differences between the methods used in the reconstruction, or to
galactic contamination in any of the data-sets.

The cross-correlation between the reconvolved data at 15 GHz
and the predictions from the COBE DMR data have been plotted as a solid
line in Fig. 2. The agreement between this cross-correlation and the
autocorrelation amplitudes  for our 15-GHz data reinforces the
conclusion that most of the signal present in our 15-GHz data
corresponds to intrinsic CMB structure.

\section{CONCLUSIONS}

We have presented new results from the Tenerife CMB experiments at 10
and 15 GHz. The sensitivity of the data at 15 GHz allows  the
detection of new individual hot and cold CMB spots. The data at 10 GHz are
noisier but sensitive enough to show evidence of CMB signals, and to
put constraints on the Galactic contamination in the data at the higher
frequency. A full joint analysis with our previous published results at
15 GHz in the largely independent strip at Dec.=+40\degg~shows the statistical consistency
between the results in both scans and puts limits on the expected
quadrupole $Q_{\rm RMS-PS}=21.0^{+6.5}_{-5.5}$ $\mu$K in the case of an
standard inflationary scenario ({\i.e.}, a Harrison-Zeldovich
spectrum).  A comparison between our results  and the COBE DMR
predictions of Bunn \et for our experiment assuming a standard CDM model shows a
clear correlation between both, and the presence of strong common features in
the region RA=$190\degg-250$\degg. In forthcoming papers we will present
in detail a separation between the CMB and the diffuse foregrounds and
will make a rigorous comparison between the Tenerife and the four-year
COBE DMR data including a joint likelihood analysis and a direct
comparison of features.

\subsection*{ACKNOWLEDGMENTS}

\noindent We thank E. F. Bunn, Y. Hoffman \& J. Silk for providing us with the results of their predictions for our experiments. The Tenerife experiments are supported by the UK Particle
Physics and Astronomy Research Council, the European Community Science
program contract SCI-ST920830, the Human Capital and Mobility
contract CHRXCT920079 and the Spanish DGICYT science program. S.
Hancock wishes to acknowledge a Research Fellowship at St. John's
College, Cambridge, UK.

\clearpage
\subsection*{FIGURE CAPTIONS}

\figcaption[fig1.eps]{The 10 and 15-GHz stacked scans at Dec=+35\degg~in the region at high Galactic latitude between RA=160\degg~and 250\degg. The data have been binned in 4\degg~bins in RA. The lower noise ($\sim 20$ $\mu$K) at 15 GHz allows the direct detection of structure at this frequency. \label{fig1}}

\figcaption[fig2.eps]{The auto-correlation $C(\theta)$ for the 15-GHz
data, showing data points and one-sigma error-bars.
The short dashed lines correspond to the uncertainties due to cosmic and sample variances for a Harrison-Zeldovich spectrum with $Q_{\rm RMS-PS}=20.1$ $\mu$K. The long-dashed lines enclose the expected 95\% confidence limits bands for
the correlation in the case of pure uncorrelated noise. The solid line
is the cross-correlation between the reconvolved results at 15 GHz and
the predictions from the COBE DMR data from Bunn \et (1996-see main text for the details). The good agreement between the correlation plots confirms the CMB origin of structure in the frequency range 15 to 90 GHz.
\label{fig2}}

\figcaption[fig3.eps]{Constraints on the quadrupole $Q_{\rm RMS-PS}$ and on the
spectral index $n$ of fluctuations obtained from a joint-likelihood
analysis of our data at Decs.=+35\degg~and +40\degg~at 15 GHz. The dashed curves show the one-sigma uncertainties. The plots indicate that for the Harrison-Zeldovich spectrum ($n=1$) $Q_{\rm RMS-PS}=21.0^{+6.5}_{-5.5}$ $\mu$K. \label{fig3}}

\figcaption[fig4.eps]{Comparison between the maximum entropy reconstruction of the Tenerife Dec.=+35\degg~data at 15 GHz (solid line) and 
the COBE DMR predictions of Bunn \et (1996) (dashed line) at 53 and 90 GHz.
\label{fig4}}

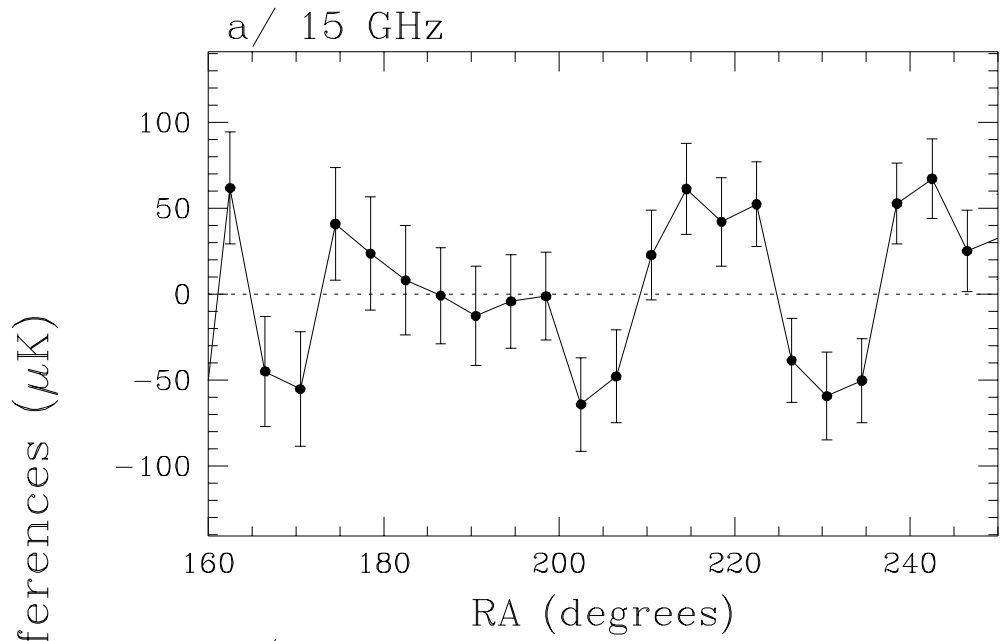
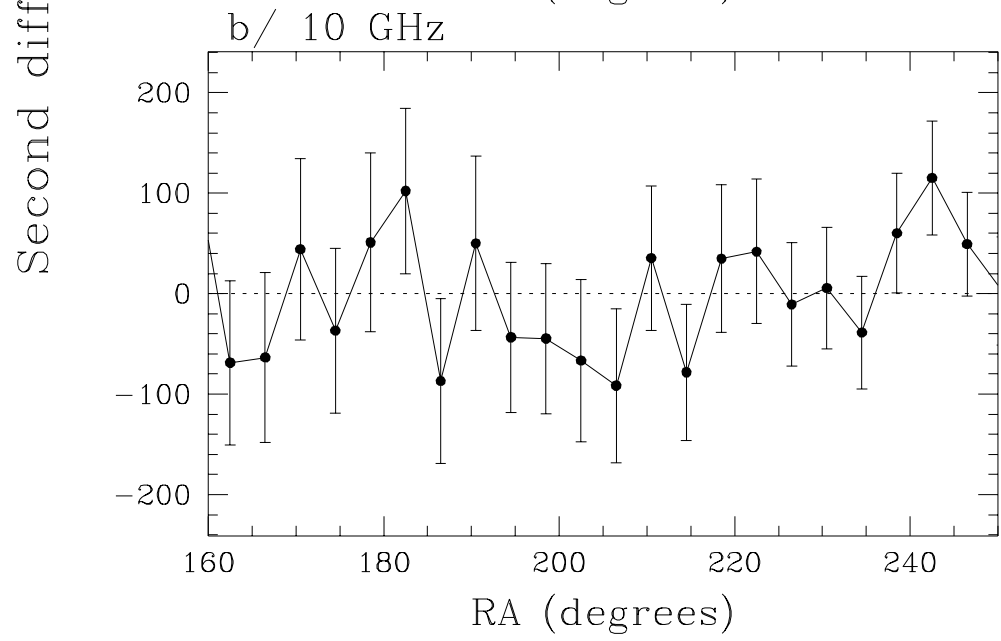
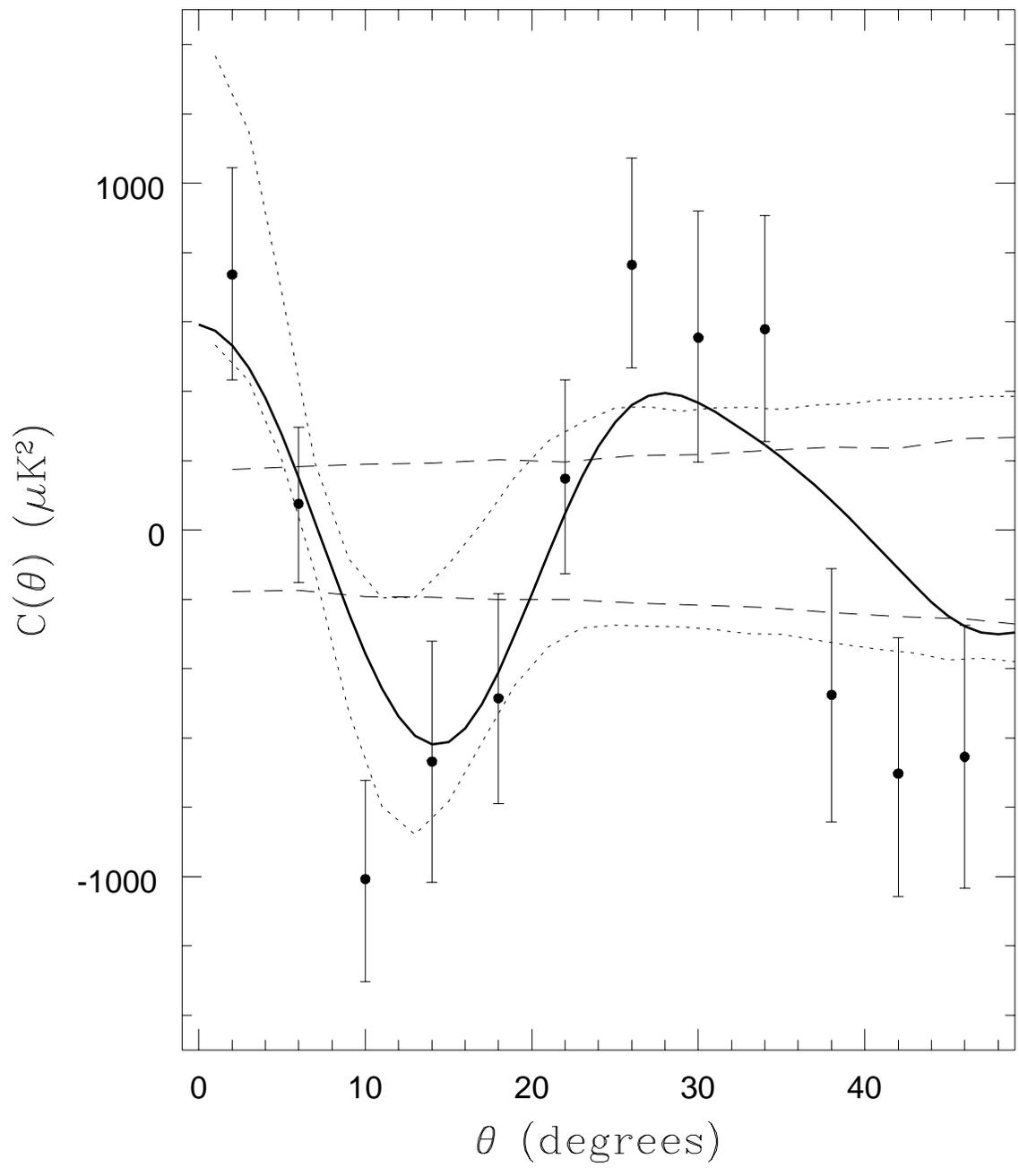
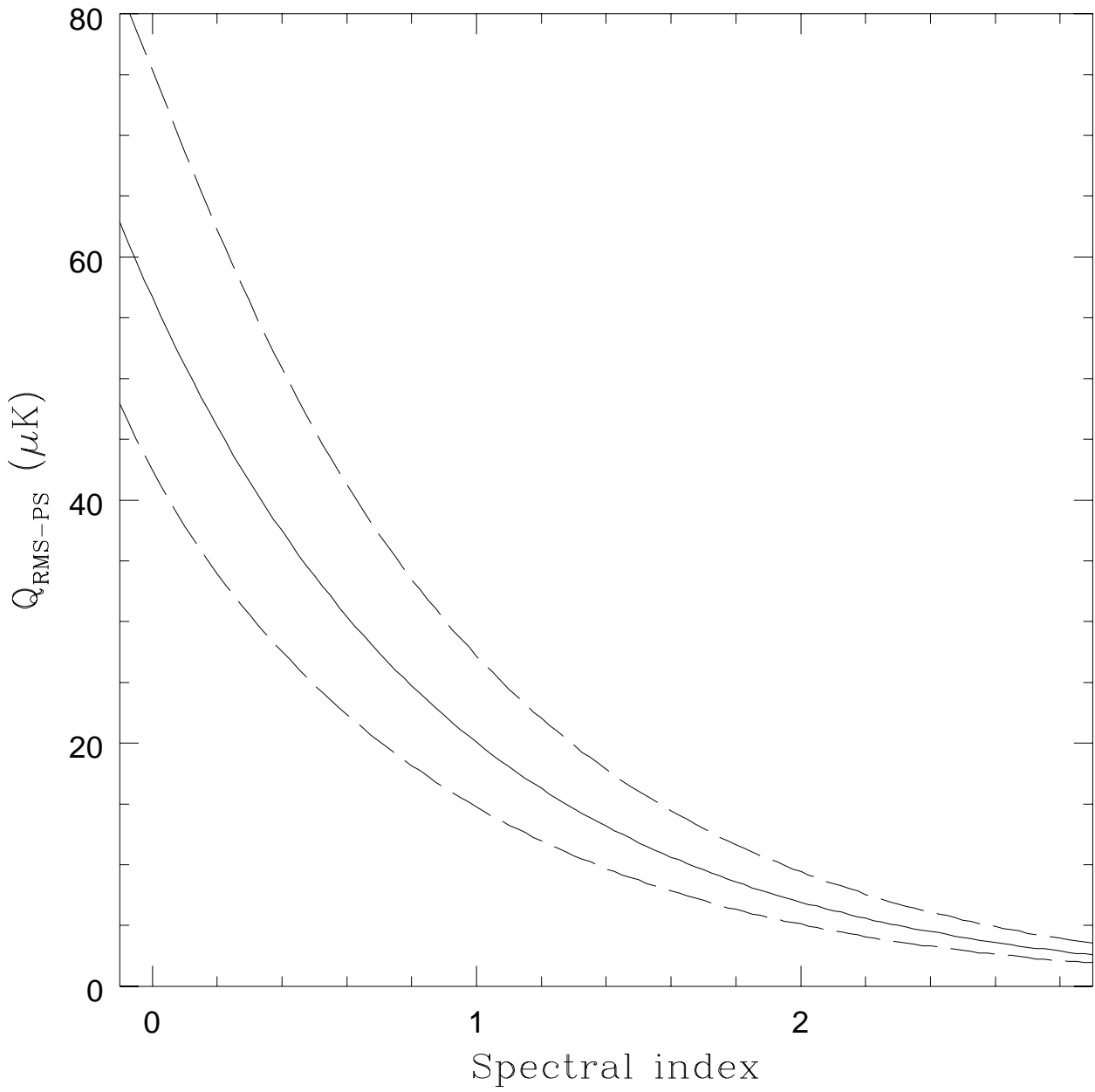
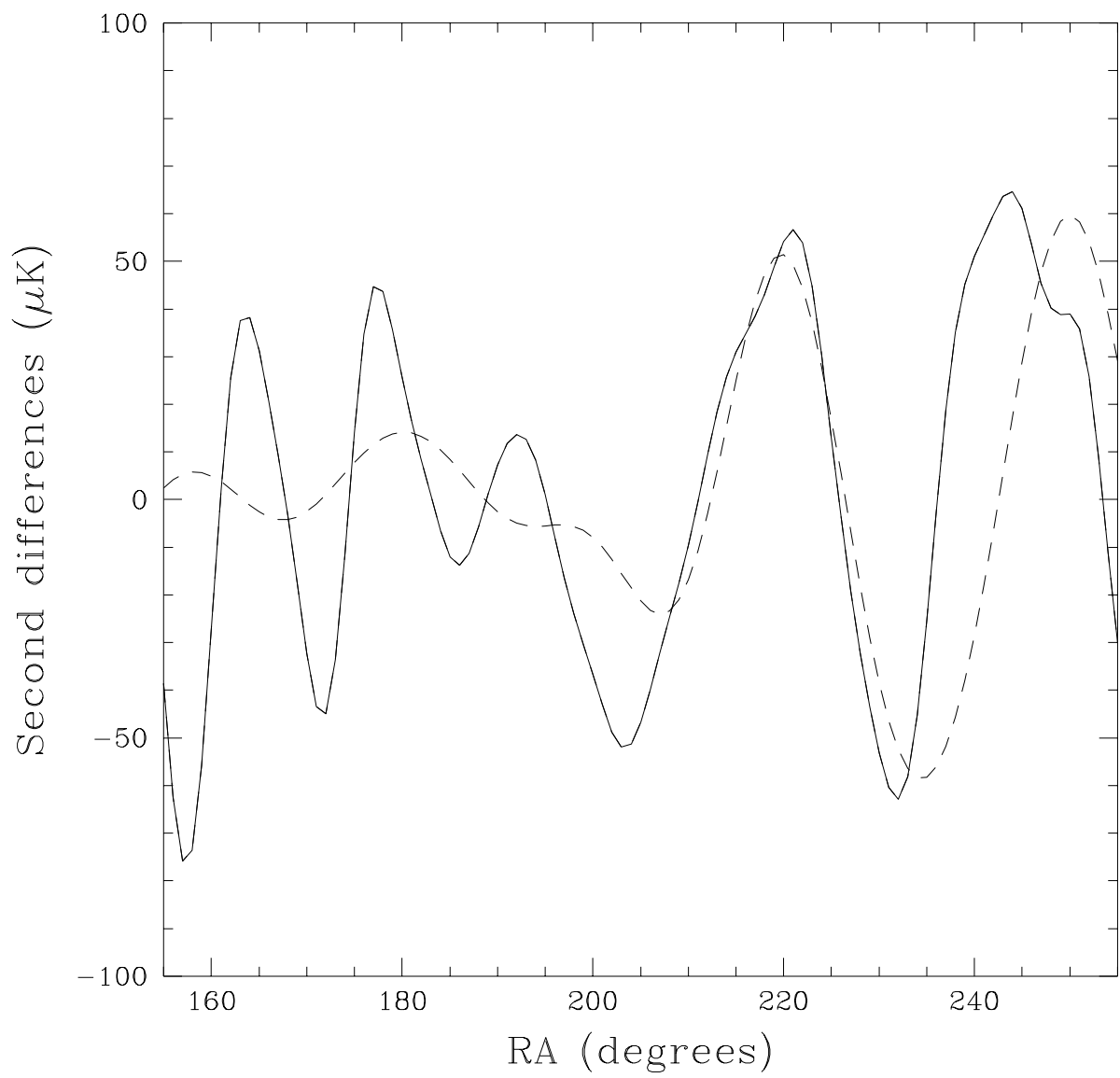
\end{document}